\documentclass[%
 reprint,
superscriptaddress,
 bibnotes,
 amsmath,amssymb,
 aps,
 prl
]{revtex4-1}

\usepackage{graphicx}
\usepackage{dcolumn}
\usepackage[title]{appendix}
\usepackage{bm}
\usepackage{braket}
\usepackage{xcolor}
\usepackage[caption=false]{subfig}
\usepackage[colorlinks=true,linkcolor=blue,citecolor=blue,filecolor=blue,urlcolor=blue]{hyperref}

\begin{document}

\preprint{APS/123-QED}

\title{Quenching of exciton recombination in strained two-dimensional monochalcogenides}

\author{J. J. Esteve-Paredes}
\email{juan.esteve@uam.es}
\affiliation{Departamento de F\'isica de la Materia Condensada, Universidad Aut\'onoma de Madrid, E-28049 Madrid, Spain.}
\author{Sahar Pakdel}
\email{pakdel.sahar@phys.au.dk}
\affiliation{Departamento de F\'isica de la Materia Condensada, Universidad Aut\'onoma de Madrid, E-28049 Madrid, Spain.}
\affiliation{Department of Physics and Astronomy, Aarhus University, 8000 Aarhus C, Denmark}
\affiliation{School of Electrical and Computer Engineering, University College of Engineering, University of Tehran, Tehran 14395-515, Iran.}
\author{J. J. Palacios}
\email{juanjose.palacios@uam.es}
\affiliation{Departamento de F\'isica de la Materia Condensada, Universidad Aut\'onoma de Madrid, E-28049 Madrid, Spain.}
\affiliation{Instituto Nicol\'as Cabrera (INC) and Condensed Matter Physics Center (IFIMAC), Universidad Aut\'onoma de Madrid, E-28049 Madrid, Spain.}
\affiliation{Department of Physics, University of Texas, Austin, Texas 78712, USA.}

\date{\today}

\begin{abstract}
 We predict that long-lived excitons with very large binding energies can also exist in a single or few layers of monochalcogenides such as GaSe. Our theoretical study shows that excitons confined by a radial local strain field are unable to recombine despite of electrons and holes co-existing in space. The localized single-particle states are calculated in the envelope function approximation based on a three-band $\boldsymbol{k}\cdot \boldsymbol{p}$ Hamiltonian obtained from DFT calculations. The binding energy and the decay rate of the exciton ground state are computed after including correlations in the basis of electron-hole pairs. The interplay between the localized strain and the caldera-type valence band, characteristic of few-layered monochalcogenides, creates localized electron and hole states with very different quantum numbers which hinders the recombination even for singlet excitons. 
 
\end{abstract}

\maketitle

{\it Introduction.---} Long-lived excitons have been long sought both for creating quantum Bose gases in semiconducting materials \cite{bose1,bose2} and for the development of novel excitonic devices. To date, these have been extensively studied in III-V and II-VI semiconductor heterostructures where the range of existence is limited to low temperatures. Very recently, they have also been observed at room temperature in van der Waals transition-metal dichalcogenide (TMD) two-dimensional (2D) heterostructures \cite{tmd1,tmd2}. In both cases the long-lived nature stems from the fact that the electron and the hole are spatially separated, thereof the name of indirect excitons.

The spatial separation between electron and hole composing an indirect or interlayer exciton in TMD 2D heterostructures allows for long recombination life-times, but it is also a limitation. Leaving aside the technical difficulties of creating heterostructures in a controlled manner, the required spatial separation limits, in turn, the binding energy. Larger inter-layer or electron-hole distances amount to longer recombination times, but weaker binding energies. Here we show that both requirements, impossible to meet in TMD heterostructures can be found in strained few-layered monochalcogenides such as GaSe. 

It is well known that the application of external strain on a semiconductor leads to a modification of the band gap. This is particularly important in 2D  crystals where large strains can be applied before the crystal rupture \cite{rupture1,rupture2}. The modulation of the band gap due to strain has been quantified for MoS$_2$ \cite{castellanosguinea}, phosphorene \cite{gabino_phosphorene}, and GaSe \cite{gabino}, to name a few. In the multilayer form of the latter and in other 2D TMDs, a causality relation between localized strain deformations and single-photon emission has been demonstrated \cite{Tonndorfgold,Tonndorf,TonndorfWse2,Kumar,Tonndorfonchip}, although the ultimate reason for this relation remains unclear.

The proven strong band-gap modulation with strain, the possibility of single-photon emission, along with an anomalous shape of the GaSe valence band, motivates the present study. Below approximately 7 layers, the valence band turns from a common inverted parabola to a caldera-type shape or ring-type shape \cite{gaseshape,gabino}. Free excitons in GaSe monolayer have been already studied with \textit{ab-initio} methods \cite{monogase}, showing a variety of behaviours due to the peculiar valence band. We are interested here in localized radial deformations as those invoked to explain single-photon emission in multilayer GaSe \cite{Tonndorfgold,Tonndorf}.  Here we show that the combined peculiarity of the valence band along with local strain quenches the photoluminescence by making the localized excitons extremely long-lived, as shown below.

\begin{figure}[t] 
	\begin{center} 
		\includegraphics[scale=0.45]{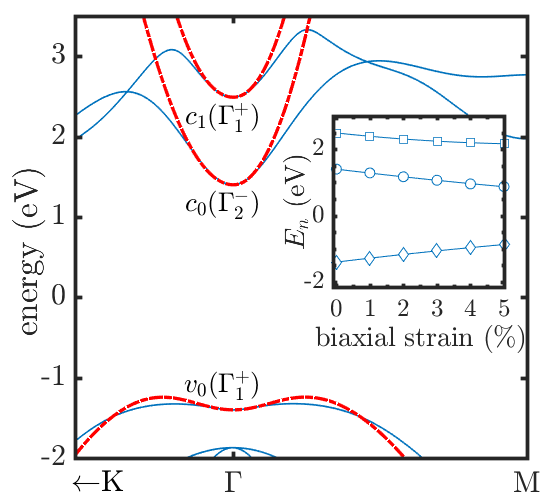}
		\caption{\small DFT (solid blue lines) and effective bands (red dashed lines) of a GaSe monolayer along the $K$-$\Gamma$-$M$ path near the $\Gamma$-point. IRs of the zone center states are shown according to Ref. \cite{Licaldera}. The inset shows the evolution of the energies at the $\Gamma$-point for several biaxial strain values, calculated using DFT.}
		\label{fig:bands}
	\end{center}
\end{figure}

{\it The effective Hamiltonian.---} We start by constructing an accurate low-energy Hamiltonian for a monolayer of GaSe using $\boldsymbol{k}\cdot \boldsymbol{p}$ theory. Firstly, we find the DFT band structure of monolayer GaSe as shown in Fig. \ref{fig:bands} (blue solid lines). We have used the CRYSTAL code  implementation of the hybrid functional HSE06 \cite{crystal14} to obtain a good approximation to the actual quasi-particle gap. Our DFT calculation will serve a two-fold purpose:  to check the validity of our effective model Hamiltonian and to obtain the necessary information (Bloch states and band edges at the $\Gamma$-point) to compute the momentum matrix elements for a parameter-free description. The unperturbed Hamiltonian is $H^{(0)}_{nm}=E_n\delta_{nm}$, where the $n,m$ indices represent Bloch states at the $\Gamma$-point, which we will use as our basis, and $E_n$ is the energy of these Bloch states. Following the  discussion in Ref. \cite{Licaldera}, we pursue a 3-band model  to properly account for the caldera form of the valence band (see Fig. \ref{fig:bands}). In $\boldsymbol{k}\cdot \boldsymbol{p}$ theory, the first order correction is linearly dependent in $k$ with the form  $H^{(1)}_{nm}=\frac{\hbar}{m_0} \sum_{\alpha}  k_{\alpha}.{p}^{\alpha}_{nm}$, where $\alpha$ (and later $\beta$) denotes Cartesian coordinates and $p^{\alpha}_{nm}$ is the $\alpha$ component of the momentum matrix element between $\Gamma$-point Bloch states. 
Based on group theory considerations we can anticipate the momentum matrix elements that are non-zero and, thereby, the general form of the Hamiltonian. We will follow the notation in Ref. \cite{Licaldera} for the $D_{3h}$ point group, to which monolayer GaSe belongs. The valence and the two first conduction bands, $v_0$, $c_0$ and $c_1$, belong to the $\Gamma^+_1$, $\Gamma^-_2$, and $\Gamma^+_1$ irreducible representations (IRs) respectively, and will form our working subspace $A$ from now on.
In order for the momentum matrix elements to be non-zero, the decomposition of the direct product of the eigenstates IRs and the IR of the momentum operators, which in the $D_{3h}$ group transform as $\Gamma^+_3$ ($p_x$, $p_y$) and $\Gamma^-_2$  ($p_z$), must contain the fully symmetrical representation $\Gamma^+_1$. One can see that only $\braket{\Gamma_3^+|p_{x,y}|\Gamma^+_1}$, $\braket{\Gamma^+_3|p_{x,y}|\Gamma^+_2}$, $\braket{\Gamma_3^-|p_{x,y}|\Gamma^-_1}$, $\braket{\Gamma^-_3|p_{x,y}|\Gamma^-_2}$, $\braket{\Gamma_3^+|p_{x,y}|\Gamma_3^+}$, $\braket{\Gamma_3^-|p_{x,y}|\Gamma_3^-}$ and their complex conjugates are non-zero. Therefore, the resulting Hamiltonian will not contain terms linear in $k$. The absence of linear terms in the Hamiltonian forces us to 
consider the coupling of the subspace $A$ with the rest of the bands, which now becomes the dominant contribution. This is typically done through the L\"owdin partitioning technique \cite{resumenkp}. Including the coupling with all bands and evaluating numerically the momentum matrix elements, the effective Hamiltonian becomes
\begin{equation} 
       H^{\text{eff}} (\boldsymbol{k})=\begin{pmatrix} A(\boldsymbol{k}) & 0 & B(\boldsymbol{k}) \\
       0 & C(\boldsymbol{k}) & 0 \\
       B^{*}(\boldsymbol{k}) & 0 &  D(\boldsymbol{k}) \end{pmatrix},
       \label{eq:Heff}
\end{equation}
with
\begin{equation}
    \begin{split}
        A(\boldsymbol{k})&=-1.4+19.6k_x^2+20.0k_y^2, \\
        B(\boldsymbol{k})&=55.5k_x^2+63.4k_y^2, \\
        C(\boldsymbol{k})&=1.4+67.7k_x^2+164.8k_y^2,\\
        D(\boldsymbol{k})&=2.5+74.3k_x^2+148.8k_y^2
    \end{split}
\end{equation}
in units of eV (and atomic units for wave vectors). As discussed earlier, $v_0$ and $c_1$ bands transform as $\Gamma_1^+$ which only couples with $\Gamma_3^+$ states. On the other hand, $\Gamma_3^+$ does not couple with $\Gamma^-_2$, the IR of $c_0$. 
As a result, one could have expected $v_0$ and $c_1$ to remain decoupled from $c_0$, as can actually be seen in Eq. \eqref{eq:Heff}. The three bands resulting from the  effective Hamiltonian are shown in Fig. \ref{fig:bands} (red dashed lines). The agreement with the original DFT bands is very good near the $\Gamma$-point, included the caldera shape of the valence band. Since we have used all the bands in the evaluation of the second order perturbation terms, no further improvement can be envision to second order within our 3-band model. 

{\it Local strain model.---} Assuming that strain changes occur on length scales much larger than the lattice spacing we can work in the envelope function framework where one writes the eigenfuntions in the form $\psi_\gamma=\sum_n F^{(\gamma)}_n(\boldsymbol{r})u_{n\boldsymbol{0}}(\boldsymbol{r})$. Here $u_{n\boldsymbol{0}}$ are the periodic Bloch functions at $\boldsymbol{k}=0$ and $F_n^{(\gamma)}$ are 2D envelope functions \footnote{Note that in the expansion of the SQD states, the envelope functions are defined in 2D while the Bloch periodic functions are \textit{ab-initio} 3D functions.} calculated from a set of coupled differential equations:
\begin{equation} \label{eq:envelope}
    \sum_{m} H_{nm}^{\text{SQD}}(\boldsymbol{r},-i\boldsymbol{\nabla}_{\boldsymbol{r}}) F^{(\gamma)}_{m}(\boldsymbol{r})=\varepsilon_\gamma F^{(\gamma)}_{n}(\boldsymbol{r}).
\end{equation}
$H^{\text{SQD}}$ is the hamiltonian of the strain-induced quantum dot (SQD) created by the local perturbation. We assume the strain to be locally biaxial which can be modeled by adding a local ``potential" to the $k$-independent diagonal terms in $H^{\text{eff}}$, 
\begin{equation} \label{eq:deltae}
    H_{nn}^{\text{SQD}}=E_{n}+\Delta_n(\boldsymbol{r})+\sum_{\alpha}M_{n}^{\alpha}\nabla^2_\alpha, 
\end{equation}
while keeping unchanged the $k$-dependent diagonal (here generically represented by $M$) and non-diagonal terms $B^{(*)}(\boldsymbol{k})$. This approximation  captures the main effect, which is the confinement of electrons and holes, while maintaining the complexity of the implementation to a minimum.  It neglects changes in the effective mass or band curvature and the possible appearance of new Hamiltonian terms due to the breaking of the $C_{3h}$ symmetry, but it has been shown to be quantitatively accurate in a somewhat similar context \cite{straincylindrical}. In practice we consider a tensile gaussian strain field of the form $S_0\exp{(-r^2/\sigma_s^2)}$ with a maximum strain at its center of 5\%.
The local lattice deformation is mapped onto $\Delta_n(r)$ thanks to a series of DFT calculations performed with biaxial tensile strain. In the inset of Fig. \ref{fig:bands} we show the $\Gamma$-point energies as obtained for several values of the tensile strain. We have also implicitly assumed, as seen in the inset of Fig. \ref{fig:bands}, that the chemical potential (set at zero) stays in the mid-gap at all points in space. A detailed calculation is beyond the scope of this work, but this choice complies with local charge neutrality  and overall thermodynamic equilibrium at low temperatures.

{\it Single-particle results.---}
\begin{figure}[t] 
	\begin{center} 
		\includegraphics[scale=0.40]{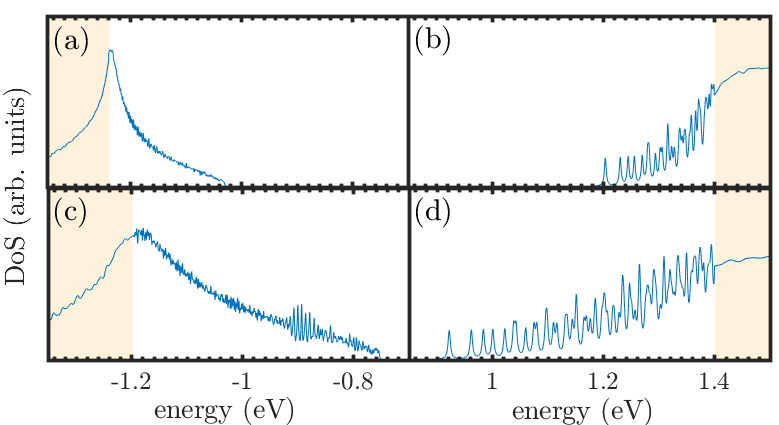}
		\caption{\small (Color online) Density of states of the valence band calculated from the numerical diagonalization of a SQD with $\sigma_s=10$ nm, $S_0=2$ \% (a) and $\sigma_s=10$ nm, $S_0=5$ \% (c).  Same for the conduction band (b and d). We have used 3600 HO basis functions for every envelope function in the two calculations with a HO frequency optimal to match the gaussian curvature of the strain field. Extended states approximately belong in the shaded regions.}
		\label{fig:density}
	\end{center}
\end{figure}
\begin{figure}[t] 
	\begin{center}
		\includegraphics[scale=0.521]{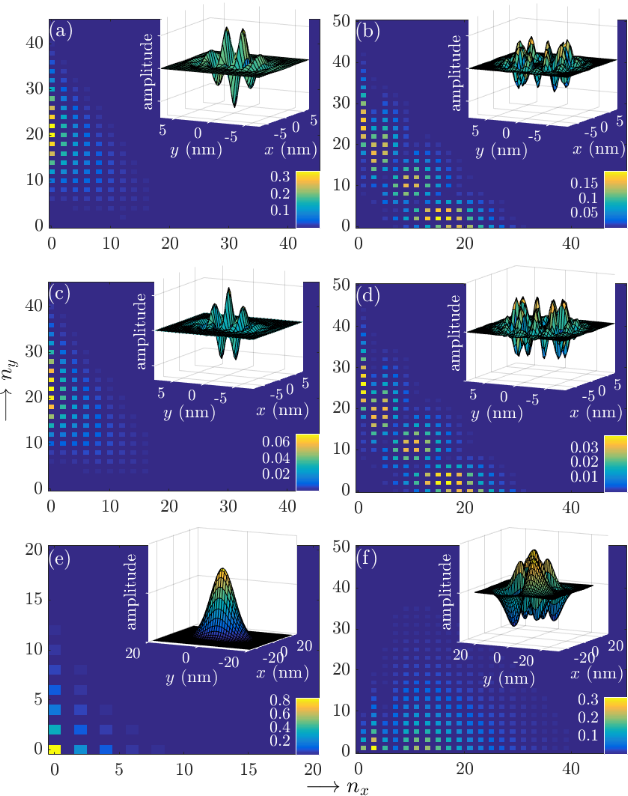}
		\caption{\small Charts of the HO expansion (absolute value of the coefficients) for the envelope functions. $v_0$ envelope function for (a) the first and (b) the 20th valence state. (c) and (d) panels show the $c_1$ envelope function for the same states (notice the much smaller contribution). $c_0$ envelope function for (e) the first and (f) the 20th conduction state. The insets show the actual shape of the associated wavefunction in real space.}
		\label{fig:hochart} 
	\end{center}
\end{figure}
In order to solve Eq. \eqref{eq:envelope} we further expand the envelope functions as
\begin{equation}
    F_n^{(\gamma)}=\sum_{n_x n_y} c^{(n \gamma)}_{n_x n_y}\phi_{n_x n_y},
\end{equation}
where $\phi_{n_x n_y}$ are the cartesian 2D harmonic oscillator (HO) basis functions. The kinetic part of the matrix elements in the HO basis can be easily evaluated using the ladder operators, while the strain contribution coming from the term $\Delta_n(r)$ is evaluated numerically. As an example we show in Fig. \ref{fig:density} the density of states of electrons and holes, obtained from the numerical diagonalization of Eq. \eqref{eq:envelope} for two representative cases. In both cases we find bound states within the band gap of the unstrained monolayer. These states distribute in a more dense spectrum in the valence band than in the conduction band, a consequence of the caldera shape of the former, which also gives rise to the expected van Hove singularity \cite{monogase}. The number of bound states in each band increases with $\sigma_s$ and $S_0$, as could intuitively be expected. The transition from bound to delocalized states (shaded regions in Fig. \ref{fig:density}) occurs near the band edges of the unstrained conduction and valence bands. A minimum of $\sim 30$ and $\sim 100$ bound states have been found in the valence and conduction band respectively in all the studied cases, which is sufficient for our purposes.

A plot of the envelope functions reveals the very different nature of valence and conduction bound states. 
In Fig. \ref{fig:hochart}(a) and \ref{fig:hochart}(c) we show a chart of the HO basis coefficients (absolute value) of the $v_0$ and $c_1$ envelope functions corresponding to the highest-energy hole state and in Fig. \ref{fig:hochart}(e) the $c_1$  envelope function of the lowest-energy electron state. Both hole envelope functions show a very similar behaviour, although $v_0$ gives the main contribution to the state. The other envelope function is zero due to the absence of coupling in Eq. \eqref{eq:Heff}. The electron state has mainly $(n_x=0,n_y=0)$ character (as could be anticipated for a parabolic band), but the hole state shows a dramatic shift to high $n_y$ quantum numbers. This behaviour can be traced back to the caldera ring away from the $\Gamma$-point in the unstrained band structure. We also show the 20th electron and hole states (moving away from the band edges) in Fig. \ref{fig:hochart}(b, d and f). The hole state spreads now also to higher values of $n_x$ while the electron state still presents the largest weight at low HO quantum numbers. 

{\it Excitons.---}  Interactions are now added to the single-particle
Hamiltonian expanded by electron and hole bound states:
\begin{equation}
    H=H^{\text{SQD}}+H^{\text{e-h}},
\end{equation}
with
\begin{equation} \label{eq: fullhamiltonian}
\begin{split}
H^{\text{SQD}}&=\sum_{i}\epsilon_{i}c_{i}^\dagger c_{i}+\sum_{j}\epsilon_{j}d_{j}^\dagger d_{j}, \\
H^{\text{e-h}} =&-\sum_{\substack{i_1,j_1,i_2,j_2 \\ \sigma, \sigma'}}c_{i_1\sigma}^\dagger c_{i_2\sigma}d_{j_2(-\sigma')}^\dagger d_{j_1(-\sigma')}V_{i_1,j_1,j_2,i_2} \\
&+\sum_{\substack{i_1,j_1,i_2,j_2 \\ \sigma, \sigma'}}S_{\sigma \sigma'} c_{i_1\sigma}^\dagger c_{i_2\sigma'}d_{j_2(-\sigma)}^\dagger d_{j_1(-\sigma')}V_{i_1,j_1,i_2,j_2},
\end{split}
\end{equation}
where $c_i$ ($d_j$) operators act on electrons (holes) and the phase factor $S_{\sigma\sigma'}=(-1)^{(1/2-\sigma-\sigma')}$ preserves essential symmetries in the electron-hole language \cite{fetterwalecka}. Note that the renormalization of the single-particle energies into quasi-particle excitations \cite{Louie} is already approximately accounted for in our hybrid DFT band structure. No other interaction terms are needed since we only consider single electron-hole pairs. At this level, excitonic states can be constructed as
\begin{equation} \label{eq:excitonexp}
    \ket{X}=\sum_{i,j,\sigma}A_{ij}c_{i\sigma}^\dagger d_{j(-\sigma)}^\dagger \ket{0},
\end{equation}
where $\ket{0}$ denotes the vacuum state with no electrons and holes. The excitonic states are then calculated by projecting Eq. \eqref{eq: fullhamiltonian}
onto the electron-hole pair basis and converged in the number of these pairs (we restrict ourselves to 200 electron-hole pairs which is enough for our purposes). Note that, to leading order, the exchange part in Eq. \eqref{eq: fullhamiltonian} will not contribute to the matrix elements \cite{Azizi,multiband}. This is due to the specific form of our Hamiltonian in Eq. \eqref{eq:Heff}, which results in conduction and valence states not sharing envelope functions. As a consequence, singlet and triplet states of the exciton are degenerate in energy. States with well defined total spin can be constructed {\it a posteriori}, if desired. 
\begin{figure}[t] 
	\begin{center} 
		\includegraphics[scale=0.30]{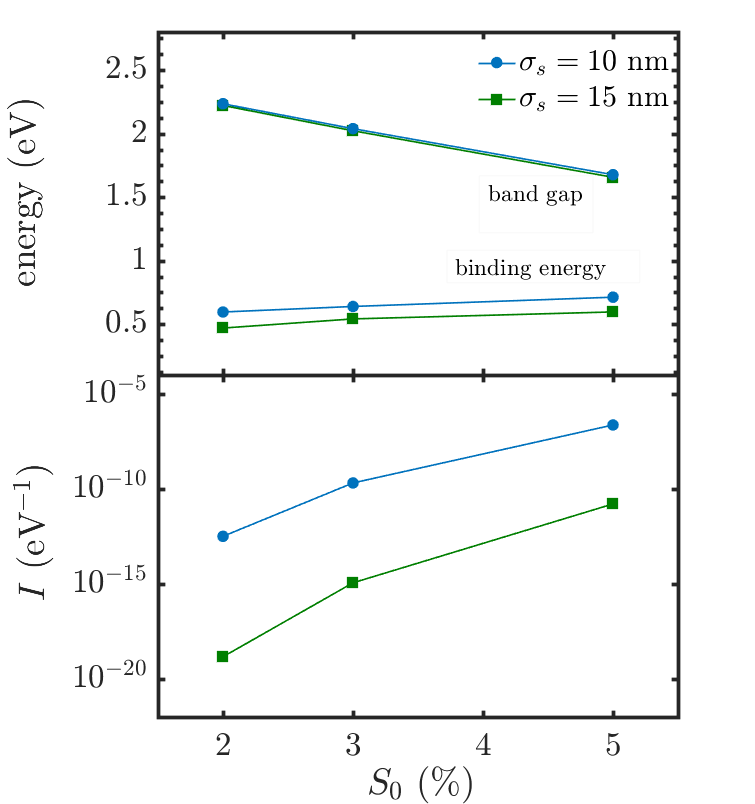}
		\caption{\small (a) Band gap and binding energy of the ground state exciton for several values of the strain strength and extension. The Keldysh interaction with a screening length $r_0=10$ \AA has been used. (b) \small Decay rate of the ground state exciton as expressed in Eq. \eqref{eq:rate} for the cases in (a).}
		\label{fig:exciton}
	\end{center}
\end{figure}
For the interaction kernel, we have considered the bare Coulomb interaction and the Keldysh interaction model, more appropriate for 2D systems \cite{keldysh}. Both types of interactions give qualitatively similar results with unimportant differences in the actual exciton binding energies which are not relevant for our discussion here. Since the interactions preserve all the symmetries, excitons can be classified according to how the transform under in-plane reflections \textit{in both} electron and hole coordinates. In the following we focus in the excitons whose electron and hole wavefunctions transform similarly, as these are, in principle, the active ones in photoluminescence.
In Fig. \ref{fig:exciton}(a) we show the quasiparticle gap and binding energy of the optically active (symmetry-allowed) lowest-energy excitonic state $E_X^{(0)}$ for different parameters characterizing our SQD. Depending on the parameters and interaction details this can be the ground state, but this is unimportant for our main conclusion (see below). In particular we have used the Keldysh potential\cite{Prada} in an environment of vacuum.
The  binding energies decrease with $\sigma_s$ since the single-particle states become more extended in space. On the other hand, the binding energy increases with the strength of the strain as they become more localized. 

{\it Quenching of the exciton recombination.---}
The most relevant result is found in the photoluminescence. The recombination rate to the ground state mediated by the emission of a photon can be computed using Fermi's golden rule (ignoring proportionality constants) \cite{Nexcitons, multiband, carbondecay}
 \begin{equation} \label{eq:emission}
    I \propto \frac{1}{E_X^{(0)}}\sum_{\boldsymbol{q}\lambda}\Big|\sum_{i,j} A_{ij} p_{ji}^{(\boldsymbol{q}\lambda)}\Big|^2\delta(E_X^{(0)}-\hbar \omega_q),
\end{equation}
Here $p_{ji}^{(\boldsymbol{q}\lambda)}$ is the scalar product of the momentum matrix element and the photon polarization vector $\boldsymbol{\epsilon}_{\boldsymbol{q}\lambda}$. The squared quantity is known as the oscillator strength. Thus, the recombination is encoded in those electron-hole pairs which have a relevant contribution to the exciton wave function. For completeness and to strengthen our conclusion, we invoke now the first order correction to the bound states \cite{resumenkp},
\begin{equation}
    \psi_{\gamma}=\sum_{n \in A}F_n^{(\gamma)}u_{n\boldsymbol{0}}+ \sum_{m \notin A}\tilde{F}_m^{(\gamma)} u_{m\boldsymbol{0}}.
\end{equation}
The first term represents the main contribution to the bound states, as explained previously, while the second one now includes Bloch functions out of the subspace $A$ of the effective Hamiltonian. The new envelope functions can be calculated as a linear combination of the derivatives of those in $A$,
\begin{equation}\label{eq:derivatives}
\tilde{F}_m^{(\gamma)}=-i\frac{\hbar}{m_0}\sum_{l}^{N_A}\frac{1}{E_l-E_m}(p_{ml}^x \partial_x + p_{ml}^y \partial_y )F_l^{(\gamma)}.
\end{equation}
With this the momentum matrix element in Eq. \eqref{eq:emission} now becomes (up to first order in $\boldsymbol{\nabla_r}$)
\begin{equation}\label{eq:momentum}
p_{j i}^{(\boldsymbol{q}\lambda)} = \sum_{m,n \in A}\bra{u_m}\boldsymbol{\epsilon}_{\boldsymbol{q}\lambda}\cdot \boldsymbol{p}\ket{u_n} \braket{F_m^{(j)} | F_n^{(i)}} 
\end{equation}
The intra-band terms (not shown), which are usually neglected \cite{multiband}, here they are strictly zero because of the structure of our Hamiltonian. For in-plane polarization, we have already seen that the momentum matrix element between states in $A$ vanish. If the polarization vector has a $z$-component, then the only coupling allowed is between $v_0$ and $c_0$ \cite{monogase}. Thus in the leading order 
\begin{equation}
    p_{j i}^{(\boldsymbol{q}\lambda)} \approx \bra{u_{v_0}}p_z\cos \theta_z \ket{u_{c_0}}\braket{F_{v_0}^{(j)}|{F_{c_0}^{(i)}}},
\end{equation}
being $\cos \theta_z$ the $z$-direction cosine of the polarization vector. For a single polarization Eq. \eqref{eq:emission} reduces to
\begin{equation} \label{eq:rate}
    I \propto \frac{|\sum_{ij}\braket{F_{v_0}^{(j)}|{F_{c_0}^{(i)}}}|^2}{E_X^{(0)}}\delta(E_X^{(0)}-\hbar \omega_q)
\end{equation}
Since we are considering symmetry-allowed active excitons, the scalar product can be non-zero. However, in our case the overlap is always vanishing small. Note again that the exciton is composed of electron-hole pairs with respective wavefunctions of very different HO quantum numbers, at least in the relevant range of energies. Therefore, at this point we can conjecture that even the dipole-allowed transitions will vanish due to the nearly orthogonal nature of the hole and electron envelope functions. In Fig. \ref{fig:exciton}(b) we show the rate of Eq. \eqref{eq:rate} in logarithmic scale. We obtain, as expected, very low values for all the cases studied. Note that in a common semiconductor with parabolic bands, one would expect the overlap between electron and hole envelope functions to be of the order of one and therefore, for similar momentum matrix elements, a decay rate many orders of magnitude larger.
Higher order corrections to the envelope functions are not expected to give a relevant contribution to Eq. \eqref{eq:momentum}. They are obtained by differentiation of the envelope functions, which translates in a spreading of the order of unity of the quantum numbers (see, e.g., Fig. \ref{fig:hochart}), thus maintaining its quasi-orthogonal character. Finally note that this result does not depend on the singlet or triplet nature of the exciton or the magnitude of the dipolar matrix elements because it is encoded in the envelope part of the wave functions. We thus conclude that photoemission is generically quenched. 

{\it Conclusions.---}
We have shown that excitons localized by a radial strain perturbation in GaSe monolayer present extremely long recombination times. This is a direct consequence of the very different nature of the wave function of electrons and holes. While the parabolic conduction band presents bound states with low HO quantum numbers, the caldera-shaped valence band makes the hole states to have a more disperse structure of HO quantum numbers. This translates in an orthogonality of electron and hole envelope functions which, in the end, quenches the recombination. Although we have focused on a specific 2D monochalcogenide, everything seems to indicate that similar results would be obtained in other monochalcogenides such as GaS and InSe \cite{gaseshape} and even in different families of materials presenting a caldera-shaped valence or conduction band.

\begin{acknowledgments}
Research supported by the Spanish MINECO through Grants FIS2016-80434-P and the Mar\'ia de Maeztu Programme for Units of Excellence in R\&D (MDM-2014-0377), the Fundaci\'on Ram\'on Areces, and the European Union Seventh Framework Programme under grant agreement No. 604391 Graphene Flagship. S.P. was also supported by the VILLUM FONDEN via the Centre of Excellence for Dirac Materials (Grant No. 11744). We acknowledge the computer resources and assistance provided by the Centro de Computaci\'on Cient\'ifica of the Universidad Aut\'onoma de Madrid.
\end{acknowledgments}

\end{document}